\documentstyle[sprocl]{article}
\input{epsf.sty}
\def\Journal#1#2#3#4{{#1} {\bf #2}, #3 (#4)}
\def\be{\begin{equation}}
\def\ee{\end{equation}}
\def\bea{\begin{eqnarray}}
\def\eea{\end{eqnarray}}
\def\note #1]{{\bf #1]}}
\def\etal{{\it et al.}}
\def\eg{{\it e.g.}}

\def\cm{\,{\rm cm}}
\def\s{\,{\rm s}}
\def\dd{{\rm d}}
\def\Fsurf{F_{\rm surf}}
\def\CK{{\cal K}}
\begin{document}
\title{HELIOSEISMOLOGY AND SOLAR NEUTRINOS}
\author{J{\o}rgen Christensen-Dalsgaard}
\address{Teoretisk Astrofysik Center, Danmarks Grundforskningsfond, and \\
Institut for Fysik og Astronomi, Aarhus Universitet, DK-8000 Aarhus C, Denmark}
\maketitle\abstracts{
Helioseismology has provided very precise information about the
solar internal sound speed and density throughout most of the solar interior.
The results are generally quite close to the properties of standard
solar models.
Since the solar oscillation frequencies do not provide direct
information about temperature and composition, 
the helioseismic results to not completely rule out an astrophysical
solution to the discrepancy between the predicted and measured
neutrino fluxes from the Sun.
However, such a solution does appear rather implausible.
}
\section{Introduction}
The persistent discrepancy between the measured fluxes of solar
neutrinos and the predictions of solar models
({\eg} Bahcall;\thinspace\cite{Bah89} for a recent review, see Bahcall, 
these proceedings)
has led to doubts about the reliability of solar-model calculations.
Indeed, a large number of suggestions have been made
for changes to the properties of the model which might reduce
the predicted neutrino fluxes.

The neutrino flux depends principally on the temperature and composition
profiles of the solar core but provide little detailed information
about the solar interior.
In the last decade far more information on the internal structure of
the Sun has been obtained from observations of solar oscillations,
providing tight constraints on solar models.\cite{GT91}
Recent ``standard'' solar models are generally in reasonable agreement 
with the helioseismic inferences.\cite{Bas96a,Bas97,Gough96,Ric96,Bah97}
Here I consider the extent to which this agreement argues against
an astrophysical solution to the solar neutrino problem.
To be definite, I base the analysis on Model S of
Christensen-Dalsgaard {\etal}\thinspace\cite{CD96}
This was computed with OPAL equation of state\thinspace\cite{Rog96}
and opacity,\cite{Igl92} using the Bahcall \& Pinsonneault\thinspace\cite{BP95}
nuclear parameters,
and including settling of helium and heavy elements
computed with the Michaud \& Proffitt\thinspace\cite{MP93} 
diffusion coefficients.
The predicted neutrino capture rates in the 
${}^{37}{\rm Cl}$ and ${}^{71}{\rm Ga}$ experiments are
8.2 and 132 SNU, respectively, while the flux of
high-energy ${}^8 {\rm B}$ neutrinos is $5.9 \times 10^6 \cm^{-2} \s^{-1}$,
quite similar to, for example, the corresponding model
of Bahcall \& Pinsonneault.\cite{BP95}

In common with other current
models of the Sun the model is based on significant simplification,
neglecting possible hydrodynamical phenomena in the convectively
stable part of the Sun, associated {\eg} with instabilities
or motion induced by convective penetration.
Indeed, it is well established that the Sun becomes unstable
to low-order, low-degree g modes during 
its evolution;\thinspace\cite{CDDG74}
the non-linear development of the instability 
might conceivable involve mixing of the solar core.\cite{DG72}
Other instabilities, leading to mixing, could be associated with
the spin-down from the commonly 
assumed early rapid rotation.\cite{Chab95}

\section{Inversion for the Solar Internal Sound Speed}
The observed solar oscillations
are adiabatic almost everywhere, to a very good approximation.
Thus their frequencies depend on 
the distribution of mass, pressure $p$ and
density $\rho$, and on the adiabatic compressibility
$\Gamma_1 = (\partial \ln p / \partial \ln \rho)_s$, the derivative
being at fixed specific entropy $s$.
Assuming that the model is spherically symmetric
and in hydrostatic equilibrium, 
the frequencies are completely determined by specifying $\rho(r)$
and $\Gamma_1(r)$, $r$ being the distance to the solar centre,
whereas they do not depend directly on temperature.
The observed modes are essentially standing acoustic waves,
with frequencies depending predominantly on the
adiabatic sound speed $c$ given by $c^2 = \Gamma_1 p/\rho$.
Departures from this simple description of the oscillations,
occurring very near the solar surface,
can be eliminated in the analysis of the frequencies.

The differences
$\delta \omega_{i} = \omega_{i}^{\rm (obs)}  - \omega_{i}^{\rm (mod)}$
between the observed and computed frequencies of the $i$-th mode
can be linearized around the model, resulting in\thinspace\cite{GThom91}
\be
\label{eq:1}
{\delta \omega_{i} \over \omega_{i}} 
= \int_0^R  K_{c^2,\rho}^{i}(r){ \delta c^2(r) \over c^2(r)}\dd r 
+ \int_0^R K_{\rho,c^2}^{i}(r) {\delta \rho(r)\over \rho(r)} \dd r 
+ {\Fsurf(\omega_{i})\over Q_{i}} + \epsilon_{i} \; .
\ee
Here the integrals extend to the surface radius $R$ of the Sun,
$\delta c^2$ and $\delta \rho$ are differences between
the Sun and the model in $c^2$ and $\rho$,
the kernels
$K_{c^2,\rho}^{i}$ and $K_{\rho,c^2}^{i}$ are known functions,
$\Fsurf(\omega_{i})$ results from the near-surface errors in the model,
and $\epsilon_{i}$ is the error in the observed frequencies.
This equation forms the basis for inferring the corrections 
to solar models.\cite{Bas96c}
The principle is to make linear combinations of eqs~(\ref{eq:1})
with coefficients $c_i(r_0)$ chosen such that the sums corresponding
to the last three terms on the right-hand side of
eqs~(\ref{eq:1}) are suppressed, while in the first term
the {\it averaging kernel}\thinspace\thinspace\thinspace\thinspace
$\CK_{c^2, \rho} (r_0, r) = \sum_i c_i(r_0) K_{c^2,\rho}^{i}(r)$
is localized in the vicinity of $r = r_0$.
The corresponding combination of the left-hand sides,
$\sum_i c_i(r_0) \delta \omega_i /\omega_i$,
then clearly provides a localized average of
$\delta c^2 / c^2$ near $r = r_0$.

\begin{figure}[ht]
\epsfxsize=9cm\hspace*{1cm}\epsfbox{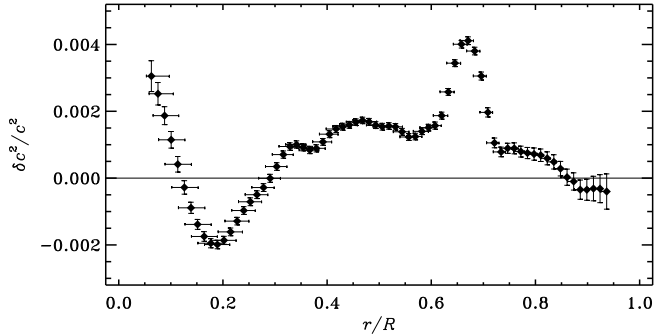}
\vskip -0.3cm
\caption{Inferred difference in squared sound speed,
in the sense (Sun) -- (model).$^4$
The horizontal error bars mark the first and third
quartile points of the averaging kernels,
whereas the vertical error bars show 1-$\sigma$ errors,
as progated from the errors in the observed frequencies.}
\end{figure}

Several extensive sets of helioseismic data are now available,
including initial results from the GONG network\thinspace\cite{Har96}
and the SOI/MDI project\thinspace\cite{Sch96} on the SOHO satellite.
The results presented here are based on a combination\thinspace\cite{Bas97}
of frequencies from the BiSON network\thinspace\cite{Els94}
and the LOWL instrument;\thinspace\cite{Tom95}
however, analyses of other, independent sets give results that 
are generally consistent with those presented here.
The inferred difference in squared sound speed between the Sun
and the model is shown in Fig.~1.
Each point corresponds to an
average of $\delta c^2/c^2$, weighted by $\CK_{c^2, \rho} (r_0, r)$;
however, as indicated these
averages are relatively well localized in $r$.
Also, the propagated data errors are small.
Thus the procedure has succeeded in providing precise and well
resolved measures of the sound-speed errors in the model,
even quite close to the centre of the Sun.
The differences are evidently highly significant; 
nonetheless, 
it is striking that the model
reproduces the solar sound speed to within a small fraction of a per cent.
This has been achieved without any explicit adjustment of parameters
to fit the model to the observations.

From the results shown in Fig.~1 we can evidently reconstruct
an estimate of the actual solar sound speed.
Tests have shown that this is largely independent of the choice
of reference model, even for models differing rather more from
the structure of the Sun than the model used here.
Thus it is possible from the observed frequencies to determine
quite precisely the dependence of sound speed on position in the Sun.

\section{Relevance to the Solar Neutrino Problem}

The support for the standard solar model provided by the helioseismic results,
at least to the level
of precision relevant to the current state of neutrino measurements,
might argue against solutions to the solar neutrino problem in
terms of non-standard solar models.\cite{Bah97}
Indeed, such models have generally been ruled out by helioseismology.
These include models with partial mixing of the core,
or with energy transport in the core 
from motion of hypothetical weakly interacting massive particles (WIMPs).
A measure of the structure of the solar core
is provided by the separation
$d_{nl} = \nu_{nl} - \nu_{n-1 \, l+2}$ in cyclic
frequencies $\nu$ between modes differing by 1 in radial order $n$
and by 2 in degree $l$, for low-degree modes.
For standard solar models the computed $d_{nl}$ is very close
to the observed value.
Core mixing increases $d_{nl}$ whereas the inclusion of WIMPs
reduces it, such that in both cases models where the neutrino
fluxes are reduced to near the measured values are inconsistent
with the observed $d_{nl}$.
Thus these models are effectively ruled out.\cite{Els90,CD91}
However, one might argue that a model combining mixing with
WIMPs could perhaps be set up in such a way as to be consistent
with both the neutrino and the oscillation data.

This exemplifies the fact that while helioseismology
constrains the mechanical properties such as sound speed and density,
other aspects of the model cannot be uniquely determined.
Indeed, 
$c^2 \propto T / \mu$,
where $T$ is temperature and $\mu$ is the mean molecular weight;
hence $T$ and $\mu$ can be varied individually, as long as
their ratio is unchanged.
For example, this might be achieved by changing
$\mu$ and $T$ through mixing and the inclusion of WIMPs, respectively.
Only by invoking the physics of stellar interiors,
such as information about energy transport and production,
is it possible to 
determine, {\eg}, the variation of 
temperature and composition,\cite{GK90,Dziem90,ST96}
and hence to compute the neutrino flux expected from the
helioseismically determined model.
Since the inferred sound speed is
so close to the standard solar model,
a substantial reduction in the neutrino flux
while keeping the model consistent with helioseismology
can be achieved only by modifying several aspects of the model computations.

Which modifications might be contemplated?
The most questionable assumption in computations of standard
models is probably the absence of motion in the solar interior;
thus it is natural to consider the possibility of mixing,
modifying the distribution of composition.
However, as argued above, any change in $\mu$ must be compensated
by changes in $T$. 
This in turn requires modification of the description of
energy generation and transport.
It is conceivable that there are significant errors in the opacities;
the basic rate of energy generation is probably rather well
known, although one cannot entirely exclude problems with 
the treatment of nuclear screening.
Evidently, inclusion of non-standard processes allows substantially
more freedom in the modelling;
examples are energy transport by WIMPs or waves, or
departures from thermal equilibrium in the present Sun,
perhaps caused by a recent mixing episode.\cite{DG72}

Models with such modifications to the 
physics permit fairly substantial reductions
in the computed neutrino flux, while keeping the sound speed
consistent with the helioseismic evidence,\cite{Rox96}
although typically drastic and perhaps unrealistic reductions 
in opacity are required.
A particularly careful analysis of this nature was carried out by
Antia \& Chitre,\cite{AC97}
who were able to reduce the computed capture rates to near the
observed values. 
However, in common with other astrophysical attempts at
a solution it was not possible to match all neutrino measurements
simultaneously.

To account for the details of the measured neutrino fluxes
Cumming \& Haxton\thinspace\cite{CH96}
proposed slow mixing of the core at such a rate that
${}^3{\rm He}$ is not in nuclear equilibrium, hence shifting
the balance between the different components of the pp chains.
This would unavoidably lead to homogenization of the hydrogen
abundance over a substantial region which, unless other modifications
were invoked, would result in a sound-speed profile inconsistent
with the helioseismic results.\cite{Bah97}
Even allowing opacity reductions by a factor of up to eight
it was not possible to obtain an acceptable sound speed.
However, further tests taking into account the modified
energy generation rate caused by the redistribution of ${}^3{\rm He}$
are still required.

\section{Conclusions}

Given the success of the standard solar model in
predicting the solar sound speed, it is tempting to
assume that the models are equally successful in determining
the temperature and composition.
Indeed, it might appear unreasonable
if the Sun were to have substantially different temperature 
and composition profiles,
arranged in such a way as to give the same sound speed as
in the standard model.
Models of this nature can be constructed, 
although only with very substantial changes to the assumed 
physics of the solar interior, carefully adjusted to avoid changes
in the sound speed.
Thus, it is perhaps natural to conclude that
the predictions of the neutrino production rates are robust,
and that therefore the neutrino discrepancies reflect a need
for revision of our understanding of the physics of the neutrino.

However, given the fundamental implications of such a conclusion,
it should not be accepted prematurely.
Further investigations, taking into account plausible errors
in the helioseismic results as well as in the physics of the
solar interior, are required to place firmer limits on
predicted neutrino fluxes consistent with the helioseismic results.

\section*{Acknowledgments}
I am grateful for permission to show the results in Fig.~1
before publication.\cite{Bas97}
This work was supported in part by
the Danish National Research Foundation through its establishment of
the Theoretical Astrophysics Center.

\end{document}